\documentclass[conference]{IEEEtran}
\IEEEoverridecommandlockouts
\usepackage{cite}
\usepackage{amsmath,amssymb,amsfonts}
\usepackage{graphicx}
\usepackage{textcomp}
\usepackage{xcolor}
\usepackage{algorithm}
\usepackage{algpseudocode}  
\algnewcommand{\Input}{\item[\textbf{Input:}]}
\algnewcommand{\Initialize}{\item[\textbf{Initialize:}]}
\algnewcommand{\Stage}[1]{\item[\textbf{#1}]}

\usepackage{amsfonts} 
\usepackage{mathrsfs} 
\usepackage{enumitem}
\usepackage{booktabs}  
\usepackage{multirow}  
\usepackage{subfigure}
\usepackage{balance} 

\usepackage{mathtools}

\def\BibTeX{{\rm B\kern-.05em{\sc i\kern-.025em b}\kern-.08em
    T\kern-.1667em\lower.7ex\hbox{E}\kern-.125emX}}
\begin{document}

\title{REG4Rec: Reasoning-Enhanced Generative Model for Large-Scale Recommendation Systems}

\author{
\IEEEauthorblockN{1\textsuperscript{st} Haibo Xing}\IEEEauthorblockA{\textit{Alibaba Group}\\
Hangzhou, China \\
xinghaibo.xhb@alibaba-inc.com}
\and
\IEEEauthorblockN{1\textsuperscript{st} Hao Deng}
\IEEEauthorblockA{\textit{Alibaba Group}\\
Beijing, China \\
denghao.deng@alibaba-inc.com}
\and
\IEEEauthorblockN{1\textsuperscript{st} Yucheng Mao}
\IEEEauthorblockA{\textit{Alibaba Group}\\
Hangzhou, China \\
maoyucheng.myc@alibaba-inc.com}
\and
\IEEEauthorblockN{2\textsuperscript{nd} Lingyu Mu}
\IEEEauthorblockA{\textit{Alibaba Group}\\
Beijing, China \\
moulingyu.mly@alibaba-inc.com}
\and
\IEEEauthorblockN{3\textsuperscript{rd} Jinxin Hu\textsuperscript{*}\thanks{* Corresponding authors}}
\IEEEauthorblockA{\textit{Alibaba Group}\\
Beijing, China \\
jinxin.hjx@alibaba-inc.com}
\and
\IEEEauthorblockN{4\textsuperscript{th} Yi Xu}
\IEEEauthorblockA{\textit{Alibaba Group}\\
Beijing, China \\
xy397404@alibaba-inc.com}
\and
\IEEEauthorblockN{5\textsuperscript{th} Hao Zhang}
\IEEEauthorblockA{\textit{Alibaba Group}\\
Beijing, China \\
zh138764@alibaba-inc.com}
\and
\IEEEauthorblockN{6\textsuperscript{th} Jiahao Wang}
\IEEEauthorblockA{\textit{Alibaba Group}\\
Beijing, China \\
wjh177423@alibaba-inc.com}
\and
\IEEEauthorblockN{7\textsuperscript{th} Shizhun Wang}
\IEEEauthorblockA{\textit{Alibaba Group}\\
Beijing, China \\
shaoan.wsz@taobao.com}
\and
\IEEEauthorblockN{8\textsuperscript{th} Yu Zhang}
\IEEEauthorblockA{\textit{Alibaba Group}\\
Beijing, China \\
daoji@alibaba-inc.com}
\and
\IEEEauthorblockN{9\textsuperscript{th} Xiaoyi Zeng}
\IEEEauthorblockA{\textit{Alibaba Group}\\
Beijing, China \\
yuanhan@taobao.com}
\and
\IEEEauthorblockN{10\textsuperscript{th} Jing Zhang\textsuperscript{*}}
\IEEEauthorblockA{\textit{Wuhan University}\\
Wuhan, China \\
jingzhang.cv@gmail.com}
}

\maketitle

\begin{abstract}
In large-scale e-commerce platforms, sequential recommendation must address diverse user behaviors driven by distinct shopping intents. Users may engage in routine purchases, flash sale shopping, cross-category exploration, or long-term decision, each requiring tailored interaction understanding. While traditional methods struggle with these complexities, generative models offer improved prediction capabilities. To better capture intent diversity in e-commerce scenarios, recent studies have introduced a reasoning process into generative recommendation.
However, these approaches are constrained by the singularity of item semantic representations, facing challenges such as limited diversity in reasoning pathways and insufficient reliability in the reasoning process. To tackle these issues, we introduce REG4Rec, a reasoning-enhanced generative model that constructs multiple dynamic semantic reasoning paths alongside a self-reflection process, ensuring high-confidence recommendations. Specifically, REG4Rec utilizes an MoE-based parallel quantization codebook (MPQ) to generate multiple unordered semantic tokens for each item, thereby constructing a larger-scale diverse reasoning space. Furthermore, to enhance the reliability of reasoning, we propose a training reasoning enhancement stage, which includes Preference Alignment for Reasoning (PARS) and a Multi-Step Reward Augmentation (MSRA) strategy. PARS uses reward functions tailored for recommendation to enhance reasoning and reflection, while MSRA introduces future multi-step actions to improve overall generalization. During inference, we further incorporate Consistency-Oriented Self-Reflection for Pruning (CORP) to correct or discard inconsistent reasoning paths, thereby mitigating noise and preventing the propagation of erroneous reasoning. Lastly, we develop an efficient offline training strategy for large-scale recommendation. 

Lastly, we develop an efficient offline training strategy tailored for large‑scale recommendation pipelines. This strategy combines dynamic embedding data management, compiler, and a layer‑adaptive dynamic quantization controller (LADQ) to reduce training time and overall hardware cost while preserving model accuracy for post‑training refinement. 
Experiments on actual datasets have verified that REG4Rec has achieved outstanding performance, realizing a performance gain of up to 16.59\%. Online experiments demonstrate its substantial practical value, including a live deployment and A/B test on Alibaba Group’s online advertising platform.
\end{abstract}


\section{Introduction}
Sequential recommendation (SR), due to its ability to predict users' next actions and uncover latent intentions by analyzing historical behavior patterns \cite{boka2024survey}, is widely applied to platforms such as e-commerce websites \cite{deng2025heterrec,lin2024enhancing}, video streaming services \cite{wang2025home}, and news portals \cite{li2022miner}. Recently, generative models have achieved significant breakthroughs in sequential recommendation \cite{zhai2024actions}. Generative Recommendation (GR) replaces traditional numeric item IDs with fixed, unique sequences of semantic tokens for item representation. By leveraging the semantic token's content representation capability and generative models' intent understanding capacity, GR dynamically generates items consistent with users' latent preferences and has quickly emerged as a new paradigm \cite{rajput2023recommender, yang2025sparse}.

\begin{figure}[t]
  \includegraphics[width=0.5\textwidth]{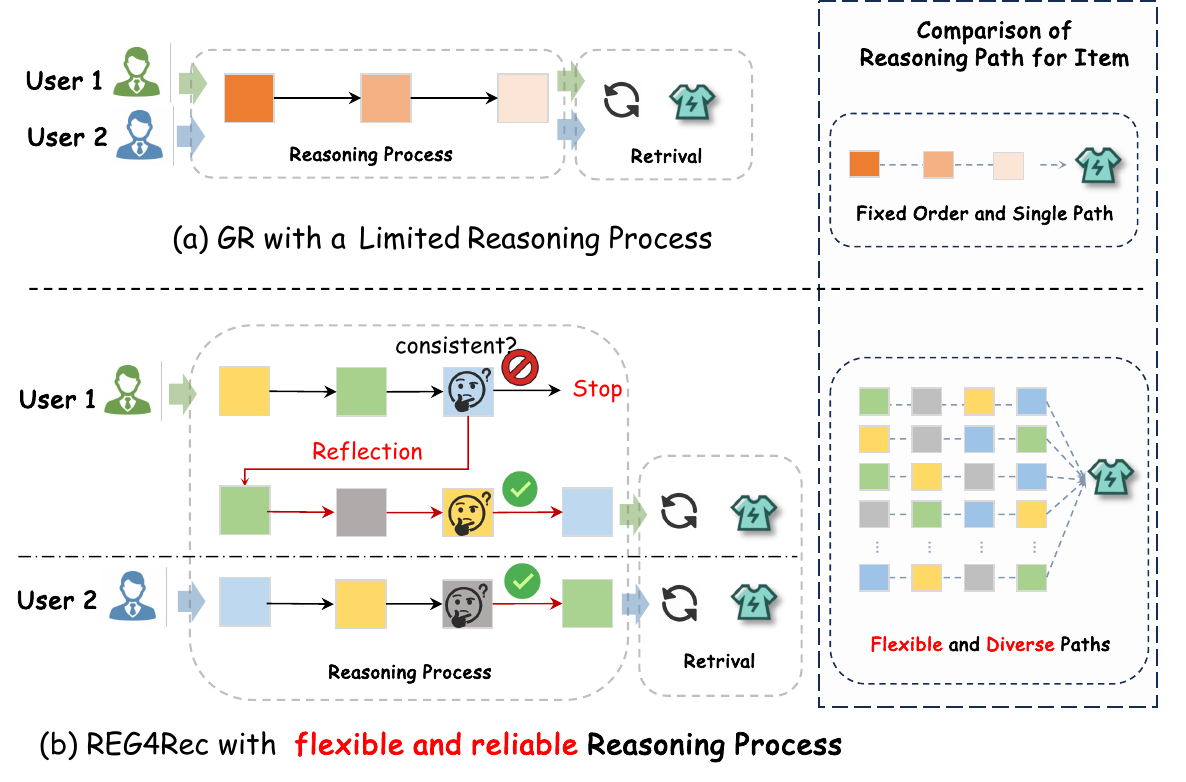}
  \caption{
  Comparison of reasoning in GR: (a) GR with a limited reasoning process; (b) our REG4Rec enables diverse and reliable reasoning process to better capture the variety of users' intents.
  }
  \label{fig:fig_intro}
\end{figure}

Recently, several studies have integrated intermediate reasoning process into GR (Figure~\ref{fig:fig_intro}(a)), generating item semantic tokens to construct paths toward final predictions and achieving significant performance gains \cite{yang2025sparse,tang2025think,zhang2025slow}. However, existing models face several limitations. 
\textbf{First}, each item is deterministically encoded into a fixed sequence of semantic tokens, producing rigid representations that conflict with the inherent uncertainty of reasoning. Since inference paths are constructed from these fixed item representations, the singularity of the representations directly results in a lack of diversity in the exploration space of reasoning pathways. Items have multi-dimensional attributes such as brand, style, and color, while different users may have diverse reasons for interest in the same item. Consequently, the constrained reasoning space of existing models fails to adequately capture this diversity of user intents, ultimately leading to suboptimal recommendation performance. \textbf{Second}, current methods lack mechanisms to assess or calibrate the reliability of reasoning paths, which may introduce uncontrolled noise during the reasoning process and reduce recommendation accuracy. 
To overcome these limitations, we aim to develop a reliable GR reasoning framework that supports diverse reasoning paths. To this end, we propose two research problems: 

\begin{itemize}[noitemsep, topsep=0pt, leftmargin=*]
\item (\textbf{RQ1}) \textbf{How can we design flexible reasoning paths to accommodate diverse user intents?} 
\item (\textbf{RQ2}) \textbf{How can we ensure a reliable reasoning process to enhance recommendation quality?}
\end{itemize}
To address these questions, we draw on core mechanisms of Large Language Models (LLMs). Their success in complex tasks stems from two advances: adopting dynamic, multi-step reasoning beyond linear thought, and incorporating self-reflective critique to improve output reliability \cite{yao2023react,yao2023tree,guo2025deepseek}. Building on these ideas, we introduce REG4Rec, a reasoning-enhanced generative recommendation model (Figure~\ref{fig:fig_intro} (b)).
To design flexible reasoning paths (RQ1), we introduce the Mixture-of-Experts (MoE)–based Parallel Quantization Codebooks (MPQ) for each item. Given multimodal inputs, MPQ uses parallel autoencoder experts to generate distinct, unordered semantic codebooks, each representing a different facet of the item. At each step, REG4Rec generates parallel predictions from all codebooks and dynamically selects the token with the highest confidence, enabling a vast combinatorial variety of paths. To ensure a reliable reasoning process (RQ2), we propose a comprehensive framework for training and inference. Specifically, we design a Preference Alignment Reward for Selection (PARS) using reinforcement learning (RL) \cite{shao2024deepseekmath} to guide the model in selecting inference paths with high consistency and confidence.
Furthermore, we propose Multi-Step Reward Augmentation (MSRA), which extends the reward horizon to future user actions, enabling better capture of long-term preferences and reducing noise from stochastic user behavior. During inference, we employ Consistency-Oriented Self-Reflection for Pruning (CORP) to identify and correct or discard inconsistent reasoning paths, thereby mitigating noise and preventing the propagation of erroneous reasoning. To make these post‑training refinements practical at industrial scale, we design an online Layer‑Adaptive Dynamic Quantization Controller (LADQ) that periodically profiles per‑layer sensitivity and assigns fp32/bf16/fp8 precision to minimize training time under an accuracy budget.

We conduct extensive experiments on four datasets (three public datasets and a large-scale industrial dataset from an online advertising system), comparing REG4Rec with six state-of-the-art baselines, such as GR models with reasoning process like ReaRec and STREAM. REG4Rec consistently outperforms all baselines across every scenario, achieving performance gains of up to 16.59\%. To validate its real-world effectiveness, we deploy REG4Rec on Alibaba’s large-scale commercial advertising platform. The subsequent online A/B test yielded a 5.60\% increase in Advertising Revenue, a 1.81\% increase in Click-Through Rate (CTR), and a 3.29\% increase in Gross Merchandise Volume (GMV). These results demonstrate the significant practical value and real-world impact of our model.

Our main contributions are summarized as follows:

\begin{itemize}[noitemsep, topsep=0pt, leftmargin=*]
\item 
We propose REG4Rec, a reasoning-enhanced generative model that constructs dynamic and reliable reasoning paths for recommendation, offering a promising solution for GR.

\item 
We design the MoE-based parallel quantization codebooks to create a flexible and diverse combinatorial space of reasoning paths. To ensure reliability, REG4Rec employs a RL-based post-training framework with Preference Alignment for Reasoning and Multi-Step Reward Augmentation to learn and select robust paths. During inference, we introduce Consistency-Oriented Self-Reflection Pruning method to remove inconsistent paths and reduce noise.

\item 
We conduct extensive experiments on both public and industrial datasets and deploy REG4Rec in an online advertising system, demonstrating its clear advantage over existing methods.
\end{itemize}

\section{Related Work}

\subsection{Sequential Recommendation}
Sequential recommendation is a fundamental task widely adopted in search, recommendation, and advertising systems  \cite{yang2025sparse,deng2025onerec,deng2025heterrec}. It predicts a user’s next item based on their chronological interaction history. To uncover latent behavioral patterns, researchers have developed various sequential modeling methods, including CNNs \cite{chen2022double,yan2019cosrec}, RNNs \cite{hidasi2015session,hidasi2018recurrent}, and Transformers \cite{kang2018self, sun2019bert4rec,pancha2022pinnerformer}. For instance, SASRec \cite{kang2018self} and BERT4Rec\cite{sun2019bert4rec} demonstrate that Transformer architectures effectively track changes in users’ interests, while PinnerFormer \cite{pancha2022pinnerformer} further extends this approach to model long-term preferences. While these models are effective, they often rely heavily on item identifiers. In large-scale recommendation scenarios, the presence of billions of distinct items can worsen the Matthew effect, resulting in poorly trained identifier parameters and restricted scalability \cite{zhai2024actions}. To address this, several works have replaced massive item-ID vocabularies with compact semantic tokens in GR models. For instance, Google’s TIGER \cite{rajput2023recommender} employs a residual quantized variational autoencoder (RQ-VAE) to create semantic tokens instead of item IDs, these tokens are then fed into a GR model to predict subsequent token IDs. Similarly, OneRec \cite{deng2025onerec} uses a residual k-means codebook to encode items and generate recommendation lists through preference alignment, while LIGER \cite{yang2024unifying} and COBRA \cite{yang2025sparse} leverages a multi-granularity GR framework that integrates semantic tokens with sparse item identifiers.

Existing GR models, which generate recommendations in a single step, often overlook the reasoning needed to capture complex intents. In contrast, we propose a reasoning-enhanced model that adaptively selects reliable reasoning paths, delivering more accurate and insightful recommendations.

\subsection{Reasoning-Enhanced Generation}
Reasoning is fundamental to solving complex problems \cite{manning2022human,wu2024large}. Recent breakthroughs in LLMs have moved beyond single-step generation, driven by two key advancements. First, LLMs create adaptive, structured reasoning flows. Instead of following a single linear path, models now explore a dynamic space of possibilities, either internally through branching thought-paths (\textit{e.g.}, Tree of Thoughts \cite{yao2023tree}) or externally by interacting with tools (\textit{e.g.}, ReAct \cite{yao2023react}). Second, they integrate reflection mechanisms for self-critique \cite{guo2025deepseek, jaech2024openai}. This involves a model drafting, evaluating, and iteratively correcting its own solutions, which significantly enhances output reliability. Inspired by these LLM advancements, researchers have begun to incorporate reasoning into recommendation systems \cite{tang2025think, zhang2025slow}. However, current approaches often implement these principles in a limited fashion. For instance, while ReaRec \cite{tang2025think} generates intermediate outcomes, its implicit nature lacks the verifiable structure required for robust reasoning. Similarly, although STREAM \cite{zhang2025slow} constructs explicit reasoning paths, its approach is overly rigid: it confines paths to textually similar items and represents each item with a fixed sequence of semantic tokens. This rigidity prevents the model from capturing the diverse intents users may have for the same item, which necessitates more adaptive reasoning.

In this paper, our method generates diverse paths within a dynamic reasoning space, while a preference alignment strategy and a reflection-based inference mechanism work in tandem to ensure the most reliable paths are selected.

\begin{figure*}[htbp]
  \includegraphics[width=1.0\textwidth]{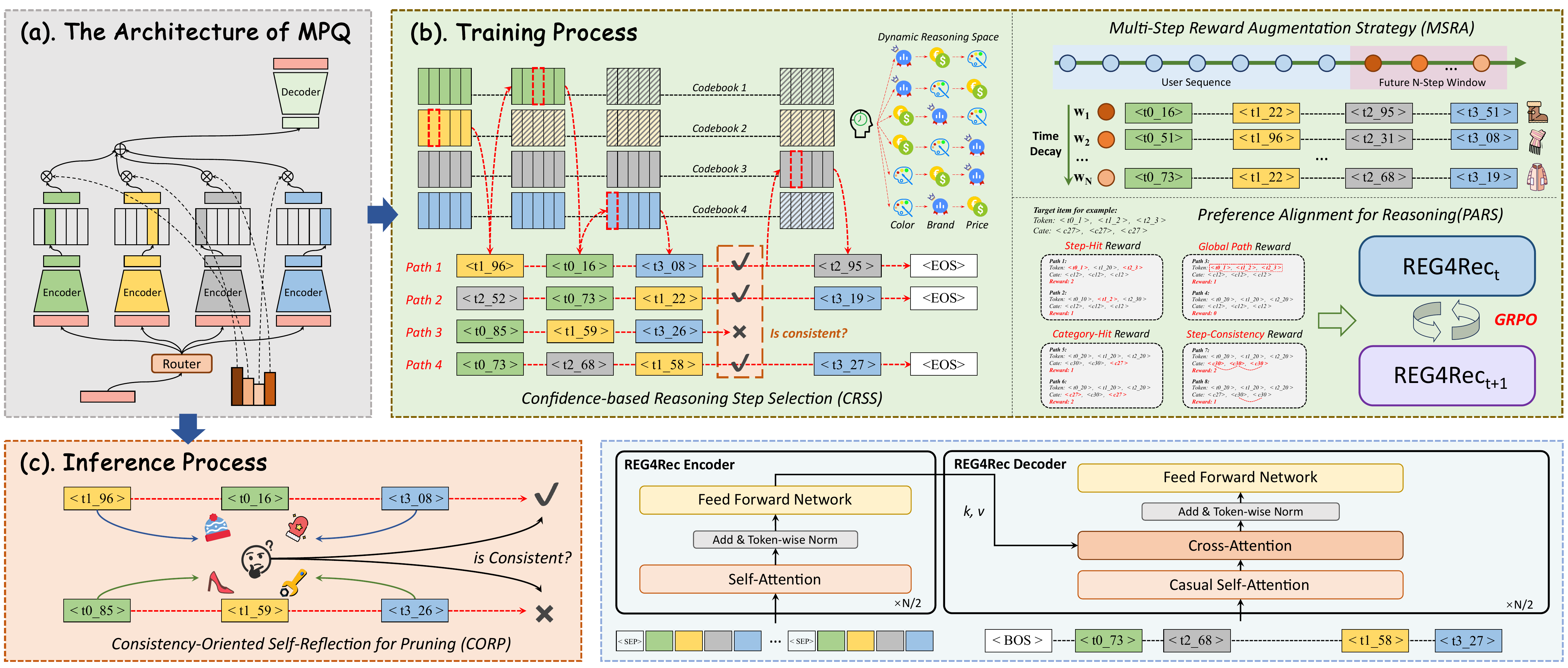}
  \caption{ 
  REG4Rec introduces MPQ for generating flexible and reliable reasoning paths. During training, CRSS diversifies path combinations, while PARS selects consistent, confident paths. MSRA extends the reward horizon to capture long-term preferences and reduce noise caused by stochastic behavior. During inference, CORP prunes inconsistent paths, preventing the propagation of errors in the reasoning process.
  }
  \label{fig:framework}
\end{figure*}

\section{Methodology}
\subsection{Problem Formulation}
This section formalizes the GR task. Let $\mathcal{I}$ denote the set of all items. For a given user $u$, the historical interaction sequence is defined as $\mathcal{S}_u = (i_1, i_2, \ldots, i_T)$, where $i_{t} \in \mathcal{I}$ is the item interacted with at time step $t$. The primary objective is to predict the user's next interaction, $i_{T+1} \in \mathcal{I}$. Inspired by prior work \cite{rajput2023recommender,yang2025sparse}, we integrate the GR model into the retrieval stage of recommender systems. A common approach is to build an encoder-decoder GR model. The model takes $\mathcal{S}_u$ as input and estimates a distribution $p(i|\mathcal{S}_u)$ over $i \in \mathcal{I}$. During inference, the model retrieves the top-$N$ items by sorting $p(i|\mathcal{S}_u)$ in descending order.

\subsection{Overview of REG4Rec}
Figure~\ref{fig:framework} illustrates the overall architecture, which consists of three main phases: codebooks construction, training, and inference.

To enable flexible reasoning, we introduce the MoE Multi-Path Quantization Codebooks (MPQ). At each step, REG4Rec treats the codebooks as parallel candidates, composing diverse reasoning paths. We also propose Confidence‑based Reasoning Step Selection (CRSS), which dynamically chooses the most confident token across codebooks at each step, adapting the path to varying user intents.

To enhance reasoning reliability, we adopt a two-stage training procedure. During pre-training, alongside predicting the target item token, we include an auxiliary category-prediction task, whose output distribution serves as a proxy for the user's primary intent and helps assess path reliability. Post-training refines reasoning via two strategies: Preference Alignment for Reasoning (PARS) and Multi-Step Reward Augmentation (MSRA). During inference, Consistency‑Oriented Self‑Reflection Pruning (CORP) evaluates step-wise consistency of reasoning paths. It prunes low-confidence paths, preventing the propagation of errors in the reasoning process.

Our method addresses \textbf{three} key objectives: 
\begin{itemize}[noitemsep, topsep=0pt, leftmargin=*]
\item 
(i) designing flexible reasoning steps to accommodate diverse user intents (Section~\ref{sec:Flexible Reasoning Path}).
\item 
(ii) enhancing reasoning reliability to improve recommendation performance (Sections~\ref{sec:Training for Reasoning Enhancement} and~\ref{sec:Inference with Reflection}).
\item 
(iii) introducing an efficient offline training strategy for large-scale recommendation (Section~\ref{sec:Efficient Training Strategy}).
\end{itemize}

\section{Flexible Reasoning Paths}
This section introduces two components that enable flexible reasoning in REG4Rec: the MoE-based parallel quantization codebooks and confidence-based reasoning step selection method. These components form the foundation of our framework.
~\label{sec:Flexible Reasoning Path}
\subsubsection{\textbf{MPQ: Moe-based Parallel Quantization Codebooks}}
At the item tokenization stage, in contrast to traditional RQ codebooks \cite{yang2025sparse}, we propose the MPQ for the reasoning process in REG4Rec. MPQ uses multiple codebooks in parallel, which are order-invariant. Each codebook captures a distinct semantic dimension of an item, yielding a factorized representation without a predefined hierarchy. Leveraging an MoE-style gating mechanism \cite{ma2018modeling}, MPQ enables information sharing across codebooks and adaptively adjusts their granularity and diversity to better capture semantics. As a result, MPQ balances the information used at each reasoning step with the overall size of the reasoning space.

Concretely, for an item $i$, its multi-modal features (\textit{e.g.}, textual, visual, and behavioral) are extracted with a pretrained model \cite{xing2025esans} and concatenated into a unified feature vector $v_{i}$. Let $M$ denote the number of parallel autoencoder experts $\{E_{r}\}^{M}_{r=1}$. Each expert $E_{r}$ is associated with a codebook $\mathcal{C}^r =\{z^r_1, z^r_2, \ldots, z^r_{D}\}$, where $z^r_j$ is a codeword and $D$ is the codebook size. Each expert operates independently, takes $v_{i}$ as input and projects it into a latent spaces $e^{r}_{i}=E_{r}(v_i)$. The experts then select the most similar codeword in $\mathcal{C}_r$ to $e^{r}_{i}$, using its index to represent a semantic token for item $i$. Formally, the index $c^i_{r}$  which is selected by the $r$-th expert represent the semantic token, is calculated as:
\begin{equation}
c^i_{r} = \arg\min_j \| e^{r}_{i} - z^r_j \|_2,
\label{Eq.1}
\end{equation}
where $\arg\min$ returns the index of the codeword with the smallest L2 distance \cite{boyd2004convex} to $e^{r}_{i}$. After obtaining $z^r_i$ from each expert, we employ a routing gate to aggregate the semantic contributions from all codebooks. The final encoding vector is $\mathbf{q}_i = \sum_{r=1}^{M}  h^{r}(v_{i})  z^{r}_{i}$, where $h^{r}(v_{i})$ denotes the routing weight assigned to expert $r$, based on the input feature vector $v_i$.

To ensure the semantic reliability and diversity of tokens in MPQ, we employ two reconstruction‑oriented loss functions. Firstly, the codebooks reconstruction loss $\mathcal{L}_{\text{recon}}$, measures how effectively the selected codewords reconstruct the input:
\begin{equation}
\mathcal{L}_{\text{recon}} = \left\| \mathbf{v}_i - \text{decoder}(\mathbf{q}_i) \right\|^2_2,
\label{Eq.3}
\end{equation}
where $\text{decoder}(\cdot)$ reconstructs an approximation of $v_{i}$ from $\mathbf{q}_i$. Additionally, we add an orthogonality regularizer on the encoder projections to promote diversity, enabling different codebooks to adaptively capture complementary dimensions of item features. Let $\{W_1, \ldots, W_M\}$ be the projection matrices for the $M$ codebooks, and $ \bar{W} $ be their column-wise L2-normalized concatenation. The orthogonality loss is:
\begin{equation}
\begin{aligned}
\mathcal{L}_{\text{orth}} &= \left\| (\bar{W}) ^\top \bar{W} - \mathbf{I} \right\|_F^2,
\end{aligned}
\label{Eq.12}
\end{equation}
where $ \mathbf{I} $ is the identity matrix and $ \|\cdot\|_F^2 $ denotes the squared Frobenius norm \cite{golub2013matrix}. The overall objective combines these terms:
\begin{equation}
\mathcal{L} =  \mathcal{L}_{\text{recon}} + \alpha  \mathcal{L}_{\text{orth}},
\label{eq:loss_function}
\end{equation}

In Eq.~\eqref{eq:loss_function}, $\alpha$ is a hyperparameter that controls the contribution of the orthogonality loss. Leveraging the MoE and parallel quantization codebooks, MPQ establishes a robust foundation for designing flexible reasoning paths.

\subsubsection{\textbf{CRSS: Confidence-based Reasoning Step Selection}}
The dynamic reasoning space constructed by MPQ utilizes multiple parallel codebooks to represent different facets of the item. To decode effectively within such a space, it is crucial to dynamically determine an ordering that balances both confidence and diversity during reasoning. To tackle this challenge, CRSS introduces a novel decoding strategy which dynamically selects the most informative token at each step using confidence scores computed from all active codebooks.

Unlike approaches where a fixed codebook is predefined for each step, CRSS performs flexible decoding by utilizing $ M $ parallel codebooks throughout the reasoning process, enabling content-aware token selection. At each decoding step, CRSS computes a predicted token and its confidence score across all $ M $ parallel codebooks. The token with the highest confidence across all codebooks is then selected as the next output, ensuring informativeness and consistency in the reasoning process.

Formally, let $ \mathcal{C_{\text{val}}} = \{ C^1, C^2, \dots, C^{K} \} $ be the active codebooks at step $j$, where $K \le M$ and $\mathcal{C_{\text{val}}} \subset \mathcal{C}$. For each $ C^k \in \mathcal{C_\text{val}} $, we compute a confidence score $ \mathbf{S}(c) $ for each token $c$ in $ C^k$. The next token $c_j$ is chosen as the one with the highest confidence over all active codebooks:
\begin{equation}
c_j = \arg\max_{C^k \in \mathcal{C_\text{val}}}( \arg\max_{c \in C^k} \mathbf{S}(c) ).
\label{Eq.13}
\end{equation}

After selecting $c_j$, its codebook is marked as used and removed from $ \mathcal{C_\text{val}} $ to avoid redundancy and maintain diversity. This process repeats with the remaining $ K - 1 $ codebooks until the active set is exhausted, yielding $ K$ tokens in total.

CRSS leverages multiple parallel codebooks in MPQ and uses confidence scores as a proxy for reliability, balancing accuracy and exploration during reasoning. This approach is well-suited to GR tasks in recommendation systems, where user behavior is highly variable and non-stationary \cite{zhang2023denoising}, and adaptive decoding is required to capture dynamic intents.
\begin{algorithm}[t]
\caption{Pre-training for Generative Recommendation}
\label{alg:training_framework}
\begin{algorithmic}[1]
\small
\Input User-item interaction data $\mathcal{D} = \{(u, \mathcal{S}_u)\}$, item feature vectors $\{\mathbf{v}_i\}_{i \in \mathcal{I}}$, number of parallel experts $M$, codebook size $D$, number of decoding steps $M$.

\Initialize Parallel autoencoder experts $\{E_r\}_{r=1}^M$, codebooks $\{\mathcal{C}^r = \{z^r_1, \dots, z^r_D\}\}_{r=1}^M$, projection matrices $\{W_r\}_{r=1}^M$, shared decoder, Transformer encoder-decoder model $\pi_\theta$, category classifier $f_c$, pre-trained feature extractors $\{\phi_t, \phi_v, \phi_b\}$.
\Stage{Stage 1: Codebook Learning}
\While{not converged}
    \ForAll{item $i \in \mathcal{I}$}
        \State $\mathbf{v}_i \gets [\phi_t(i), \phi_v(i), \phi_b(i)]$
        \For{$r = 1$ to $M$}
            \State $e^r_i \gets E_r(\mathbf{v}_i)$
            \State $c^i_r \gets \arg\min_j \| e^r_i - z^r_j \|_2$
            \State $z^r_i \gets z^r_{c^i_r}$ 
        \EndFor
        
        \State Compute routing weights: $h^r(\mathbf{v}_i) \gets \text{MoE-gate}(\mathbf{v}_i)$
        \State $\mathbf{q}_i \gets \sum_{r=1}^{M} h^r(\mathbf{v}_i) \cdot z^r_i$ 
        
        \State $\mathcal{L}_{\text{recon}} \gets \left\| \mathbf{v}_i - \text{decoder}(\mathbf{q}_i) \right\|_2^2$ 
        
        \State Normalize matrices: $\bar{W} \gets \text{L2-norm}([W_1, \dots, W_M])$
        \State $\mathcal{L}_{\text{orth}} \gets \left\| \bar{W}^\top \bar{W} - \mathbf{I} \right\|_F^2$ 
        
        \State $\mathcal{L} \gets \mathcal{L}_{\text{recon}} + \alpha \cdot \mathcal{L}_{\text{orth}}$ 
        
        \State Update $\{E_r , \mathcal{C}^r, W_r,  \pi_\theta\}$ via gradient descent on $\mathcal{L}$
    \EndFor
\EndWhile
\Stage{Stage 2: Pre-training}
\While{not converged}
    \State Sample $(u, \mathcal{S}_u) \sim \mathcal{D}$
    \State Encode history: $H \gets \text{Encoder}(\mathcal{S}_u)$
    \State Decode token sequence $y = [c_1, \dots, c_M]$
    \State Compute token loss: $\mathcal{L}_{\text{token}} \gets -\sum_{i=1}^{M} \log p(c_i \mid c_{<i}, H)$
    \State Predict categories: $g_i^{\text{gr}} \gets f_c(c_{<i}, H),\ \forall i$
    \State Compute category loss: $\mathcal{L}_{\text{cate}} \gets -\log \tilde{p}(g_i^{\text{gr}} \mid c_{<i}, H)$
    \State Update $\pi_\theta$ via gradient descent on $\mathcal{L}_{\text{token}} + \lambda \mathcal{L}_{\text{cate}}$
\EndWhile

\end{algorithmic}
\end{algorithm}

\section{Training for Reliable Reasoning Process}
~\label{sec:Training for Reasoning Enhancement}
We adopt a two‑stage training framework: pre-training and post-training. In pretraining, in addition to predicting the target item token, we add an auxiliary category‑prediction task to support subsequent reliability assessment. In post-training, to improve the reliability and quality of the generated reasoning paths, we propose two complementary strategies: Preference Alignment for Reasoning and Multi-Step Reward Augmentation Strategy.

\subsubsection{\textbf{Pretrain task}}
During pre-training, the encoder processes the token sets from a user’s historical items, and the decoder predicts the token set of the next item. Formally, the training objective aims to optimize the following probability:
\begin{equation}
\mathcal{L}_{\text{token}} = -\sum_{c=1}^{M} \log p(c^{T+1}_i \mid c^{T+1}_{<i}, \mathcal{S}_u),
\end{equation}
where $c^{T+1}_i$ denotes the $i$-th token of the next item, and $c^{T+1}_{<i}$ are previously generated tokens. To encourage consistency and improve reasoning accuracy, we add an auxiliary category prediction at each decoding step:
\begin{equation}
\mathcal{L}_{\text{cate}} = -\log \tilde{p}(g_i^{\text{gr}}\mid c^{T+1}_{<i}, \mathcal{S}_u),
\end{equation}
where $ g_i^{\text{gr}} $ is the predicted category of the $ i $-th decoding step.

\subsubsection{\textbf{PARS: Preference Alignment for Reasoning}} 
In this section, we introduce a RL framework designed to optimize reasoning paths. PARS frames item recommendation as a sequential decision-making: at each step, REF4Rec selects a token from MPQ’s parallel codebooks while exploring multiple potential paths to discover optimal reasoning paths. The objective is to learn a policy that generates reliable, semantically coherent reasoning paths aligned with user preferences.

Let $\pi_\theta(y | \mathcal{S}_u)$ be the policy parameterized by $\theta$, which generates a tokens sequence $y = \{c_1, c_2, \ldots, c_{M}\}$ conditioned on the user's history $\mathcal{S}_u$. During training, we collect a set of paths $\tau = \{y_1, y_2, \ldots, y_G\}$ sampled from earlier versions of the current policy. Here, $G$ represents the number of paths, which are used to stabilize policy updates over time. These paths are used to estimate advantages $A_i$, which measure how well each generated path $y_i$ performs relative to others. The advantage function is defined as follows:
\begin{equation}
\begin{aligned}
A_i = \frac{r_i - \text{mean}(\{r_1, r_2, ..., r_G\})}{\text{std}(\{r_1, r_2, ..., r_G\})},
\end{aligned}
\end{equation}
where $r_i$ is the cumulative reward of path $y_i$. Following GRPO algorithm \cite{shao2024deepseekmath}, which employs policy optimization with Kullback-Leibler (KL) divergence \cite{kullback1951information} for stability, we define the objective function as follows:
\begin{equation}
\begin{aligned}
\begin{split}
J(\theta) &= \mathbb{E}_{s_u \sim D, \{y_i\}_{i=1}^G \sim \pi_{\theta_{\text{old}}}(\cdot|s_u)} \\
&\quad \left[ \frac{1}{G} \sum_{i=1}^G \min \left( \frac{\pi_\theta(y_i | s_u)}{\pi_{\theta_{\text{old}}}(y_i | s_u)} A_i, \right. \right. \\
&\quad\quad \left. \left. \text{clip} \left( \frac{\pi_\theta(y_i | s_u)}{\pi_{\theta_{\text{old}}}(y_i | s_u)}, 1 - \epsilon, 1 + \epsilon \right) A_i \right) \right. \\
&\quad\quad \left. - \beta D_{\text{KL}}(\pi_\theta \| \pi_{\theta_{\text{ref}}}) \right] ,
\end{split}
\end{aligned}
\label{Eq.grpo}
\end{equation}
where $\pi_{\theta_{\text{old}}}$ denotes the reference policy, $D_{\text{KL}}(\cdot||\cdot)$ represents the KL divergence measuring the discrepancy between the current policy and the reference policy.

To further enhance the quality of the generated reasoning paths, we design \textbf{four reward functions} to improve semantic consistency, diversity, and alignment with ground-truth patterns.

\textbf{Step-Hit Reward ($ \gamma_{\text{step}} $)}: This reward measures the proportion of generated tokens that belong to the target item’s ground-truth token set, ensuring semantic alignment between the reasoning path and the target item's representation. Each item is associated with a predefined set of $ M $ tokens. We define the reward as:
\begin{equation}
\gamma_{\text{step}}=\frac{1}{M}\sum_{i=1}^{M}\mathbb{I}(c_i^{\text{gr}} \in \mathcal{C}^i),
\end{equation}
where $M$ denotes the number of reasoning steps, $ c_i^{\text{gr}} $ is the $ i $-th token generated by REG4Rec, and $\mathcal{C}^i$ is the ground-truth token set of the target item. The indicator function $ \mathbb{I}(\cdot) $ equals 1 if the generated token is in the ground-truth set and 0 otherwise. This reward treats the ground-truth token set as unordered, aggregating matches across all generated tokens. The token set of item in MPQ is unordered. This design encourages the model to explore diverse reasoning paths while maintaining semantic alignment.

\textbf{Category-Hit Reward ($ \gamma_{\text{cate}} $)}: At each step $i$, REG4Rec predicts a category probability distribution via a pretrained classifier. The predicted category $g_i^{\text{gr}}$ at step $i$ is compared with the target item’s ground-truth label $ g^{\text{gt}} $. This evaluation serving as a metric for assessing how well the generated reasoning path remains aligned with the item's ground-truth semantics across all steps, ensuring semantic consistency throughout the entire reasoning process. Formally, the category-hit reward is defined as:
\begin{equation}
\gamma_{\text{cate}} = \frac{1}{M} \sum_{i=1}^{M} \mathbb{I}(g_i^{\text{gr}} = g_i^{\text{gt}}),
\end{equation}
where the indicator function $\mathbb{I}(\cdot)$ returns 1 if $g_i^{\text{gr}}$ matches $g_i^{\text{gt}}$, and 0 otherwise. This reward evaluates semantic consistency between predicted and ground-truth categories across reasoning steps, promoting alignment with high-level item semantics.

\textbf{Step-Consistency Reward ($ \gamma_{\text{js}} $)}: To promote semantic coherence across consecutive steps, we calculate the Jensen-Shannon (JS) divergence \cite{lin2002divergence} between the predicted category distributions $\mathcal{G}$ at steps $i$ and $i+1$:
\begin{equation}
JS(\tilde{p}_i \| \tilde{p}_{i+1}) = \frac{1}{2} D_{\text{KL}}(\tilde{p}_i \| m) + \frac{1}{2} D_{\text{KL}}(\tilde{p}_{i+1} \| m),
\end{equation}
where $m = \frac{1}{2} (\tilde{p}_i + \tilde{p}_{i+1})$, and $\tilde{p}_i$ denotes the predicted probability distribution over categories at step $i$, while $\tilde{p}_{i+1}$ represents the corresponding distribution at step $i+1$. Here, $D_{\text{KL}}(\cdot \| \cdot)$ denotes the Kullback-Leibler (KL) divergence, which measures the dissimilarity between two probability distributions.

To measure the overall consistency across the reasoning path, we aggregate the JS divergence values over all adjacent distribution pairs to compute a final step-consistency reward:
\begin{equation}
\gamma_{\text{js}} = 1 - \frac{1}{M-1} \sum_{i=1}^{M-1} JS(\tilde{p}_i \| \tilde{p}_{i+1}).
\label{Eq.step_js}
\end{equation}

In Eq.~\eqref{Eq.step_js}, by subtracting the above value from 1, the reward $\gamma_{\text{js}}$ increases as consecutive category distributions become more similar, promoting smooth transitions between reasoning steps.

\textbf{Global Path Reward ($ \gamma_{\text{path}} $)}: We evaluate path robustness with a progressive retrieval scheme that allows up to $d$ tokens to be incorrect, testing the model's ability to maintain retrieval performance under partial errors. For a reasoning path of length $M$, we build a cumulative sum of codebook representation $z$ using only the $M-d$ correct tokens, $z=\sum_{m=1}^{M-d} z^{m}$, and perform top-$N$ retrieval from the item pool to check whether the target item is still retrieved. The reward under $d$ errors is:
\begin{equation}
\gamma_{\text{path}} =
\begin{cases}
1 & \mathbb{I}(i_{T+1} \in \text{Top-}N(z)), \\
0 & \text{otherwise}.
\end{cases}
\label{Eq.step_global_path}
\end{equation}
In Eq.~\eqref{Eq.step_global_path}, the indicator function $\mathbb{I}(\cdot)$ return 1 if the target item is ranked within the top-$N$ retrieved candidates and 0 otherwise. 

The overall reward for post-training is the sum of the above components, enabling jointly optimization of semantic alignment, path-level robustness, and step-wise consistency. Rather than restricting learning to a single ground-truth path, PARS encourages exploration of diverse reasoning path while ensuring semantic consistency and accuracy, improving adaptability and performance in recommendation tasks.

\subsubsection{\textbf{MSRA: Multi-Step Reward Augmentation Strategy}}
Traditional reward designs align the model with a single next item $ i_{T+1} $. However, this strict one-step alignment often fails to capture the complexities of real-world sequential behaviors \cite{deng2025heterrec}. In real-world recommendation system, user intent emerges over short horizons and is noisy and dynamic. To address these challenges, we propose a MSRA Strategy that expands valid targets to future interactions within a short time window.

MSRA treat the next $ h $ items $ i_{t+1}, i_{t+2}, \dots, i_{t+h} $ as valid representations of current interest. This captures broader temporal dependencies, allowing the model to explore diverse reasoning paths and generalize to varied user behaviors. Rather than strictly aligning with a single ground-truth item, we evaluate the generated path against all items in the $h$ window and compute step-hit reward ($ \gamma_{\text{step}} $) and category-hit reward ($ \gamma_{\text{cate}} $).

To reflect the natural decline in the influence over time, we apply exponential decay when aggregating rewards. For each item $ i_{t+j} $ ($ j = 1, \dots, h $), we assign an exponential weight $ w^{j-1} $, where the time-decay coefficient $ w \in (0, 1) $ determines how quickly the influence diminishes as $ j $ increases. The final reward is computed as the normalized weighted average, ensuring that rewards are appropriately scaled and aggregated across the time window. The $\gamma_{\text{step}}^{\text{msra}}$ is defined as:
\begin{equation}
\gamma_{\text{step}}^{\text{msra}} = \frac{\sum_{j=1}^{h} w^{j-1}  \gamma_{\text{step}}^{(j)}}{\sum_{j=1}^{h} w^{j-1}},
\end{equation}
where $ \gamma_{\text{step}}^{(j)} $ denotes the step-hit reward of the generated reasoning path with respect to item $ i_{t+j} $. The category‑hit reward $ \gamma_{\text{cate}}^{\text{msra}} $ is defined analogously by replacing $ \gamma_{\text{step}} $ with $ \gamma_{\text{cate}} $.

Additionally, We also extend the global path reward. Rather than requiring exact retrieval of the single target $ i_{t+1} $, the augmented path reward is positive if the top-$N$ results include any of the $ h $ candidate items $\{i_{t+1}, \dots, i_{t+h} \}$. This relaxation encourages the model to accommodate broader user intents. Formally, let $ W_t = \{i_{t+1}, \dots, i_{t+h}\} $ represent the set of $ h $ candidate items, and let $ \text{Top-}N(z) $ be the top-$N$ retrieved items using the cumulative representation $ z $ of the generated path:
\begin{equation}
\gamma_{\text{path}}^{\text{msra}} = \frac{\sum_{j=1}^{h} \left( w^{j-1}  \mathbb{I}(i_{t+j} \in \text{Top-}N(z)) \right)}{\sum_{j=1}^{h} w^{j-1}},
\end{equation}
where $ \mathbb{I}(\cdot) $ denotes an indicator function that returns 1 if any item in the candidate set is included in the top-$N$ results and 0 otherwise. By applying time-decayed weighting, MSRA prioritizes recent interactions while preserving sensitivity to long-term preferences, improving the robustness and diversity of the reasoning paths.



\begin{algorithm}[t]
\caption{Post-training for Generative Reasoning}
\label{alg:post_training}
\begin{algorithmic}[1]
\Input User-item sequences $\mathcal{D} = \{(u, \mathcal{S}_u)\}$, MPQ codebooks $\{\mathcal{C}^r\}$, decoding steps $M$, rollout number $G$, future window $h$, time-decay factor $w \in (0,1)$, clip range $\epsilon$, KL weight $\beta$

\Initialize Policy $\pi_\theta$, old policy $\pi_{\theta_{\text{old}}} \gets \pi_\theta$, reference policy $\pi_{\theta_{\text{ref}}} \gets \pi_\theta$, category classifier $f_c$

\While{not converged}
    \State Sample $(u, \mathcal{S}_u) \sim \mathcal{D}$, encode history $H \gets \text{Encoder}(\mathcal{S}_u)$
    \State Sample $G$ rollouts: $\tau = \{y_1, \dots, y_G\} \sim \pi_{\theta_{\text{old}}}(\cdot \mid H)$
    
    \ForAll{$y_i = [c_1, \dots, c_M] \in \tau$}
        \State Predict categories: $g_k^{\text{gr}} \gets f_c(c_{<k}, H),\ \forall k$
        \State Candidate items: $W = \{i_{t+1}, \dots, i_{t+h}\}$
        \State $Z \gets \sum_{j=1}^{h} w^{j-1}$, initialize $\gamma^{\text{msra}} \gets 0$
        
        \For{$j = 1$ to $h$}
            \State $\gamma_{\text{step}}^{(j)} \gets \frac{1}{M} \sum_{k=1}^{M} \mathbb{I}(c_k \in \mathcal{C}^k(i_{t+j}))$
            \State $\gamma_{\text{cate}}^{(j)} \gets \frac{1}{M} \sum_{k=1}^{M} \mathbb{I}(g_k^{\text{gr}} = g^{\text{gt}}(i_{t+j}))$
            \State $z \gets \sum_{k=1}^{M} z_k$, $\mathbb{I}_{\text{path}}^{(j)} \gets \mathbb{I}(i_{t+j} \in \text{Top-}N(z))$
            \State $\gamma^{\text{msra}} \mathrel{+}= w^{j-1} \cdot \left( \gamma_{\text{step}}^{(j)} + \gamma_{\text{cate}}^{(j)} + \mathbb{I}_{\text{path}}^{(j)} \right)$
        \EndFor
        
        \State Normalize MSRA rewards: $\gamma^{\text{msra}} \gets \gamma^{\text{msra}} / Z$
        \State $\gamma_{\text{js}} \gets 1 - \frac{1}{M-1} \sum_{k=1}^{M-1} JS(\tilde{p}_k \| \tilde{p}_{k+1})$
        \State Total reward: $r_i \gets \gamma^{\text{msra}} + \gamma_{\text{js}}$
    \EndFor
    
    \State Normalize advantages: $A_i \gets (r_i - \mu_r) / \sigma_r$
    \State Compute ratio: $\rho_i \gets \pi_\theta(y_i \mid H) / \pi_{\theta_{\text{old}}}(y_i \mid H)$
    \State Update $\pi_\theta$ by gradient ascent on via Equation \ref{Eq.grpo}, then $\pi_{\theta_{\text{old}}} \gets \pi_\theta$
\EndWhile
\end{algorithmic}
\end{algorithm}
\section{Deployment of REG4Rec}
\subsection{Efficient Training Strategy}
~\label{sec:Efficient Training Strategy}
\begin{figure}[htbp]
  \includegraphics[width=0.48\textwidth]{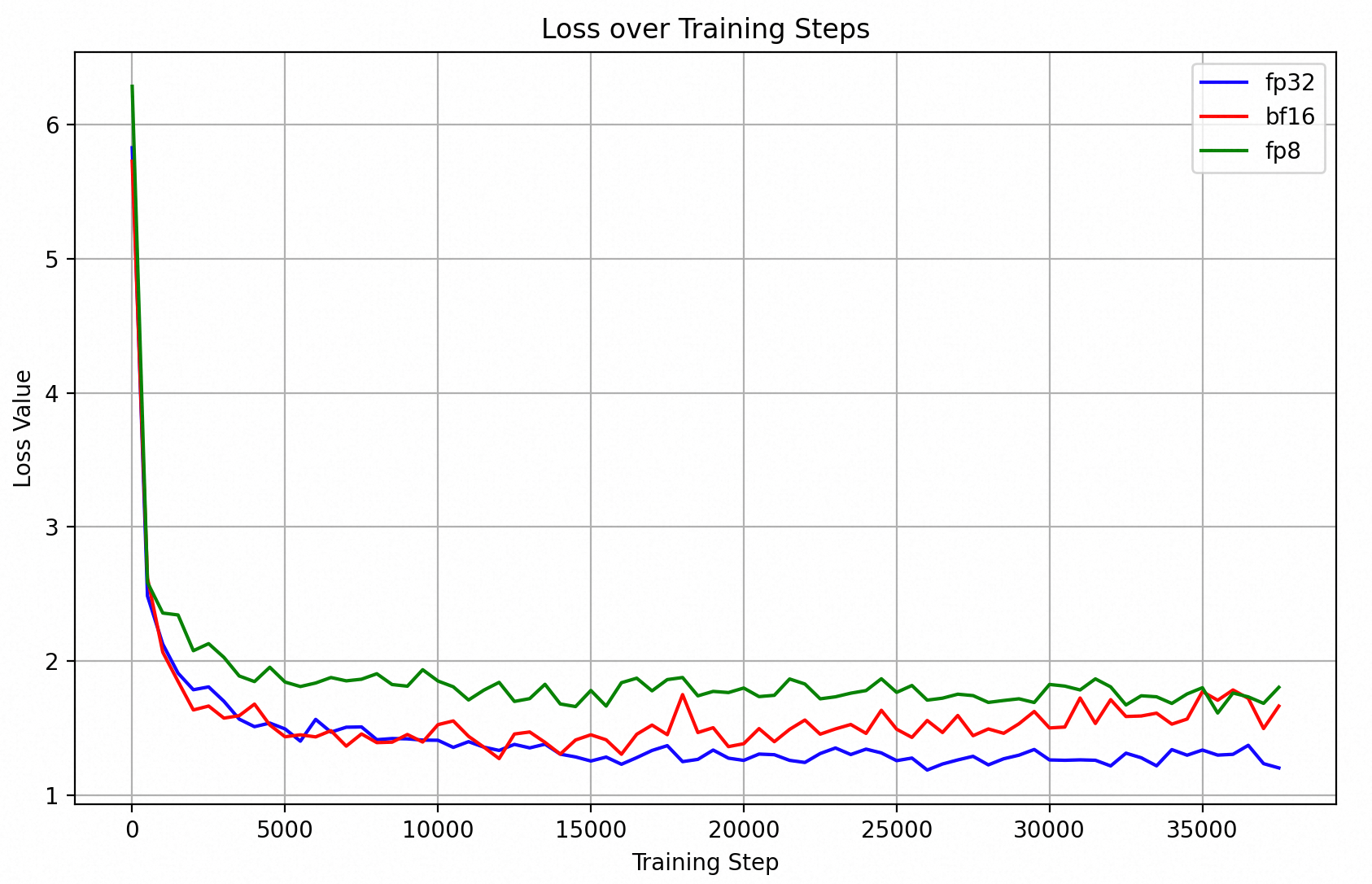}
  \caption{
  Loss progression across precisions.
  }
  \label{fig:loss_steps}
\end{figure}

While post-training significantly improves model performance, it introduces substantial offline training costs, a critical concern in large-scale recommender systems. A common acceleration strategy is to reduce numerical precision. However, conventional pipelines typically adopt a uniform, fixed format for all layers, such as fp32 (full-precision floating point), bf16 (half-precision floating point), or fp8 (an 8-bit floating‑point format), ignoring the inherent heterogeneity and temporal dynamics of different model components \cite{rouhani2023microscaling,li2024contemporary,wang2018training}. Our initial attempts to apply standard mixed-precision training (AMP) with BF16 resulted in noticeable accuracy degradation compared to the FP32 baseline. As shown in Figure \ref{fig:loss_steps}, the loss trajectories for BF16 and FP8 plateaued prematurely in later training stages, while FP32 continued to converge.

Through investigation, we identified two root causes for this performance drop: (i) Quantization Sensitivity: Activations and gradients in certain layers are highly sensitive to low precision, frequently underflowing to zero in BF16 and causing information loss. (ii) Insufficient Gradient Scaling: The standard gradient scaling in AMP was inadequate to compensate for the extremely small gradients in these sensitive layers. This motivated a more intelligent, fine-grained approach to quantization.

To balance computational efficiency and model accuracy, we propose the Layer-Adaptive Dynamic Quantization Controller (LADQ). The core principle of LADQ is to dynamically allocate higher precision to quantization-sensitive layers while assigning lower precision to robust, computationally-intensive layers. To implement this, we first needed a reliable method to measure layer-wise sensitivity during training. We explored several candidate metrics traditionally used for post-training quantization, including the Hessian matrix \cite{dong2020hawq}, the Fisher matrix, and the K-FAC matrix \cite{martens2015optimizing}, which are conventionally applied for quantization sensitivity measurement during inference. We extended these approaches through sampling-based measurement during the training process.

\begin{figure}[htbp]
  \includegraphics[width=0.48\textwidth]{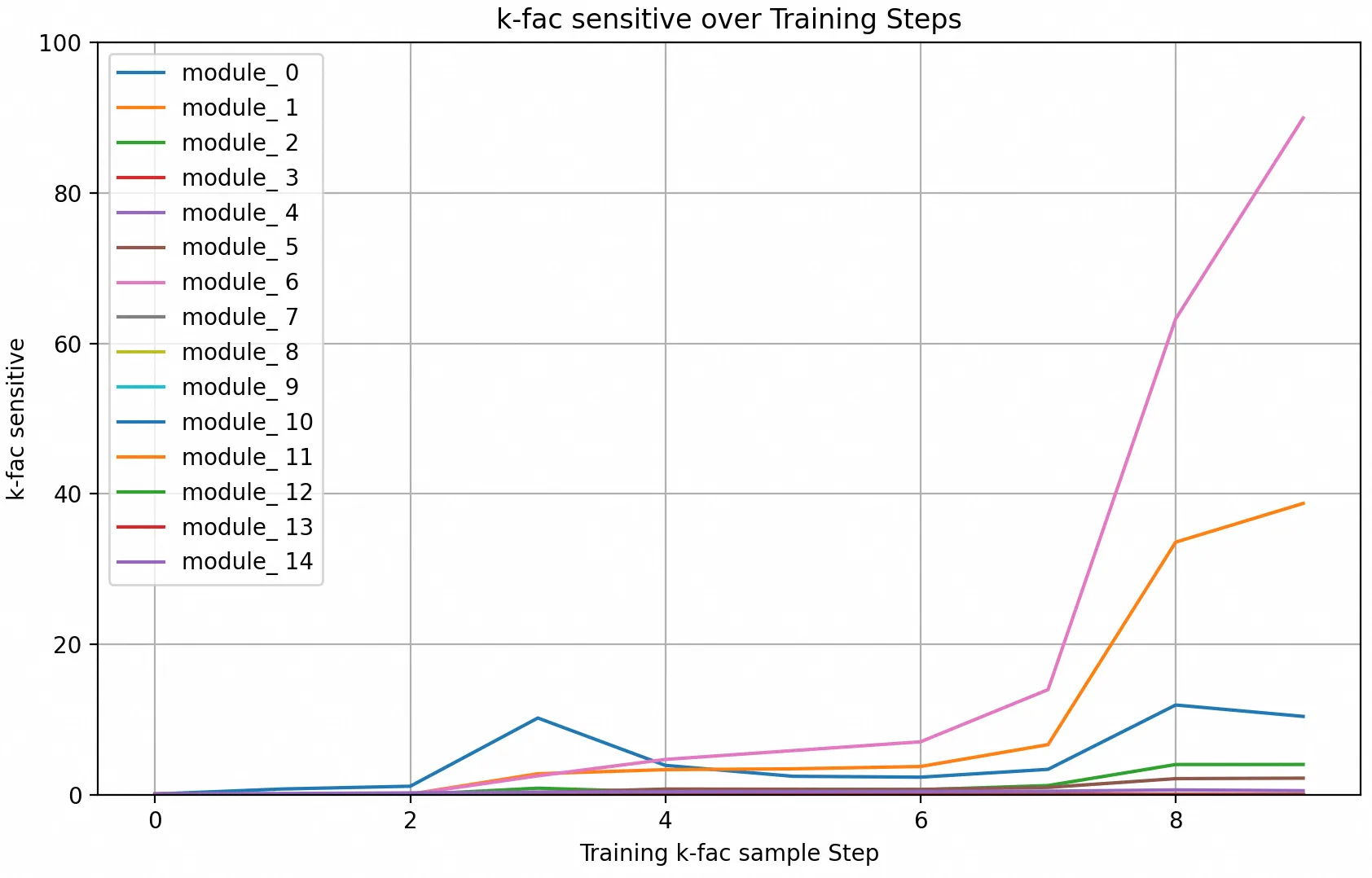}
  \caption{
  K-FAC matrix trace sampling during training.
  }
  \label{fig:KFAC}
\end{figure}

Through repeated sampling during training, we derived two key findings that inform LADQ's design:
\begin{itemize}[noitemsep, topsep=0pt, leftmargin=*]
\item \textbf{Dynamic Distribution}: The set of quantization sensitive layers is not static but shifts across different training phases.
\item \textbf{Hybrid Sensitivity}: Layer-wise sensitivity varies significantly. For example, as illustrated in Figure \ref{fig:KFAC}, within our model's transformer architecture, module 6 (the attention mechanism in the second layer) exhibited pronounced sensitivity, particularly during late-stage training.
\end{itemize}

To validate sensitivity metrics, we manually tested quantization configurations. Results confirmed that preserving high precision for sensitive modules (\textit{e.g.}, module 6) minimized loss degradation. Among Hessian, Fisher, and K-FAC, K-FAC proved optimal due to its superior accuracy-computation tradeoff.

Based on these insights, we designed the LADQ controller to automate precision allocation. This controller operates periodically throughout the training process, executing a three-step cycle:
\begin{itemize}[noitemsep, topsep=0pt, leftmargin=*]
\item  (i). Computes per-layer sensitivity.
\item  (ii). Calculates a reward score for each layer, which jointly considers sensitivity and contribution to overall training latency.
\item (iii). Dynamically assigns the optimal precision from the set \{fp32, bf16, fp8\} to each layer, prioritizing higher precision for high-reward (sensitive but critical) layers.
\end{itemize}

By adapting precision at a layer-wise level based on dynamic sensitivity, LADQ successfully accelerates the training phase while maintaining loss convergence on par with the full-precision baseline, providing a practical solution for resource-constrained environments.

\subsection{Inference with Reflection}
~\label{sec:Inference with Reflection}
Inspired by the consistency-oriented rewards in our GRPO framework, we propose the Consistency Oriented Self‑Reflection Pruning Strategy (CORP). This approach refines the standard beam search by incorporating a measure of reasoning consistency into the path selection process, thereby steering the generation towards more logically coherent sequences and mitigating the introduction of noise.

The core mechanism is as follows. During beam search, for each candidate path being extended, we do not only consider the probability of the next token. At every $s$-th step, the category predictor produces a probability distribution $\tilde{p}_i$ over item categories, serving as a semantic signal for consistency checking. This distribution acts as a snapshot of the model's understanding or reasoning focus at this step.

Specifically, consistency is measured by JS divergence between consecutive distributions (Eq.~\eqref{Eq.step_js}). If $JS(\tilde{p}_i \| \tilde{p}_{i+s})$ exceeds the self-reflection threshold $\theta$, it signals potential inconsistency or low confidence. The model then triggers self-reflection, terminating the current path and rolling back to an earlier step for regeneration. This pruning preserves only semantically coherent and logically consistent paths, enhancing overall recommendation accuracy.

\section{Offline Experiments}
In this section, we conduct extensive experiments on four datasets to demonstrate the performance of our proposed method and address the following research questions:
\begin{itemize}[noitemsep, topsep=0pt, leftmargin=*]
\item \textbf{RQ1}: How does REG4Rec perform in comparison with other state-of-the-art models?
\item \textbf{RQ2}: How does each component contribute to the overall performance of REG4Rec?
\item \textbf{RQ3}: How do hyperparameters influence the model performance?
\item \textbf{RQ4}: How does performance scale with the number of reasoning steps at inference?
\item \textbf{RQ5}: How effective is our proposed training strategy in REG4Rec, LADQ, in balancing training efficiency and model accuracy?
\item \textbf{RQ6:} Can REG4Rec deliver performance improvements in real-world online recommendation platforms?

\end{itemize}
\begin{table}[htbp]
    \centering
    \caption{Statistics of Public and Industrial Datasets. \#Users and \#Items denote the number of users and items, respectively; \#Interactions denotes the total number of user–item interactions in the dataset.}
    \renewcommand{\arraystretch}{1.3}
    \label{fig:data_scale_comparison}
    \resizebox{240pt}{!}{  
        \begin{tabular}{c|c|c|c|c}
            \toprule
            Dataset & Beauty & Sports & Toys & Industrial Dataset \\
            \midrule
            \#User & 22,363 & 35,598 & 35,598 & 35,154,135 \\
            \#Item & 12,101 & 18,357 & 18,357 & 48,106,880 \\
            \#Interaction & 198,360 & 296,175 & 167,526 & 5,730,321,793 \\
            \bottomrule
        \end{tabular}
    }
\end{table}
\subsection{Experimental Setup}
\textbf{Dataset.} Four datasets are used to demonstrate the performance of our proposed method. The first three are public datasets, while the fourth one is an in-house industrial dataset. Table~\ref{fig:data_scale_comparison} summarizes dataset statistics.
\begin{itemize}[noitemsep, topsep=0pt, leftmargin=*]
\item \textbf{Amazon Product Reviews.} We use three publicly available benchmarks from the Amazon Product Reviews dataset \cite{AmazonDataset}, which contains user reviews and item meta information from May 1996 to July 2014. For GR, we focus on three categories: "Beauty", "Sports and Outdoors", and "Toys and Games". We generate training sequences from users’ review histories, sorted by timestamp, and exclude users with fewer than five reviews to ensure sufficient interactions. We follow the standard leave-one-out evaluation protocol \cite{kang2018self}: the last item is used for testing, the second-to-last for validation, and all preceding items for training.

\item \textbf{Industrial Dataset.} We construct the in-house offline dataset from sequential user behaviors and feedback logs collected on the Alibaba’s e-commerce advertising platform. The dataset spans diverse recommendation scenarios, including homepage recommendations, product detail pages, and interactive gaming sections, capturing varied user behaviors. It contains data from 30 million users and 40 million advertisements collected between October 2024 and May 2025, offering a comprehensive view of real-world user behavior and ad content.
\end{itemize}

\begin{table*}[htbp!]
\centering
\caption{Performance Comparison across baselines. The best metric for each (dataset, metric) pair is highlighted in bold, while the second-best is underlined. The last column (Improv.) denotes the relative improvement of our REG4Rec over the best baseline.}
\renewcommand{\arraystretch}{1.3}
\resizebox{\linewidth}{!}{
\begin{tabular}{cc|ccccccc|c}
\toprule
\textbf{Dataset} & \textbf{Metric} & \textbf{SASRec} & \textbf{$S^3$-Rec} & \textbf{TIGER} & \textbf{COBRA} & \textbf{ReaRec} & \textbf{STREAM} & \textbf{REG4Rec} & \textbf{Improv.} \\
\midrule
\multirow{4}{*}{Beauty} 
& R@5  & 0.0387 & 0.0387 & 0.0454 & 0.0537 & 0.0431 & \underline{0.0538} & \textbf{0.0586} & +8.92\% \\
& N@5  & 0.0249 & 0.0244 & 0.0321 & \underline{0.0395} & 0.0276 & 0.0388 & \textbf{0.0416} & +5.32\% \\
& R@10 & 0.0605 & 0.0647 & 0.0648 & \underline{0.0725} & 0.0654 & 0.0691 & \textbf{0.0792} & +9.24\% \\
& N@10 & 0.0318 & 0.0327 & 0.0384 & \underline{0.0456} & 0.0375 & 0.0422 & \textbf{0.0493} & +8.11\% \\
\midrule
\multirow{4}{*}{Sports} 
& R@5  & 0.0233 & 0.0251 & 0.0264 & 0.0305 & 0.0256 & \underline{0.0324} & \textbf{0.0372} & +14.81\% \\
& N@5  & 0.0154 & 0.0161 & 0.0181 & 0.0215 & 0.0179 & \underline{0.0232} & \textbf{0.0252} & +8.62\% \\
& R@10 & 0.0350 & 0.0385 & 0.0400 & 0.0434 & 0.0393 & \underline{0.0452} & \textbf{0.0527} & +16.59\% \\
& N@10 & 0.0192 & 0.0204 & 0.0225 & 0.0257 & 0.0219 & \underline{0.0271} & \textbf{0.0296} & +9.23\% \\
\midrule
\multirow{4}{*}{Toys} 
& R@5  & 0.0463 & 0.0443 & 0.0521 & 0.0619 & 0.0552 & \underline{0.0644} & \textbf{0.0676} & +4.97\% \\
& N@5  & 0.0306 & 0.0294 & 0.0371 & \underline{0.0462} & 0.0374 & 0.0442 & \textbf{0.0493} & +6.71\% \\
& R@10 & 0.0675 & 0.0700 & 0.0712 & 0.0781 & 0.0747 & \underline{0.0799} & \textbf{0.0862} & +7.88\% \\
& N@10 & 0.0374 & 0.0376 & 0.0432 & \underline{0.0515} & 0.0435 & 0.0502 & \textbf{0.0574} & +11.46\% \\
\midrule
\multirow{4}{*}{Industrial Dataset} 
& R@5  & 0.0658 & 0.0759 & 0.0903 & 0.0963 & 0.0742 & \underline{0.0981} & \textbf{0.1094} & +11.52\% \\
& N@5  & 0.0453 & 0.0530 & 0.0594 & 0.0646 & 0.0517 & \underline{0.0680} & \textbf{0.0738} & +8.53\% \\
& R@10 & 0.0916 & 0.1066 & 0.1293 & 0.1397 & 0.0982 & \underline{0.1422} & \textbf{0.1569} & +10.34\% \\
& N@10 & 0.0564 & 0.0628 & 0.0715 & 0.0785 & 0.0603 & \underline{0.0804} & \textbf{0.0890} & +10.70\% \\
\bottomrule
\end{tabular}
}
\label{table:offline_experiment_results}
\end{table*}

\noindent \textbf{Baselines.}
To evaluate the performance of our method, we compare REG4Rec against \textbf{six representative baselines}. These baselines are described as follows:
\begin{itemize}[noitemsep, topsep=0pt, leftmargin=*]
\item \textbf{SASRec \cite{kang2018self}}\ : A transformer-based model that employs the self-attention mechanism to capture long-term dependencies.
\item \textbf{$S^3$-Rec \cite{zhou2020s3}}\ : A transformer-based model that integrates self-supervised learning to enhance representation learning.
\item \textbf{TIGER \cite{rajput2023recommender}}\ : A GR model built on an encoder-decoder structure, leveraging semantic IDs (based on RQ-VAE) and autoregressive decoder to improve the generalizable of recommended items.
\item \textbf{COBRA \cite{yang2025sparse}}\ : A GR model with a decoder-only structure, incorporating semantic IDs (based on RQ-VAE) and dense vectors to capture semantic insights and collaborative signals.
\item \textbf{ReaRec \cite{tang2025think}}\ : A transformer-based model that extends SASRec to generate multiple implicit intermediate outcomes without explicitly defined inference paths.
\item \textbf{STREAM \cite{zhang2025slow}}\ : A transformer-based model uses textually similar items as explicit reasoning steps, guided by the GRPO algorithm for reasoning.
\end{itemize}

\noindent \textbf{Evaluation Metrics.}
For evaluation, we adopt the Recall@K (R@K) and NDCG@K (N@K) as standard metrics for recommendation, following \cite{yang2025sparse}, with $K$ set to 5 and 10. These metrics assess both the accuracy of item retrieval and the quality of the ranking.

\noindent\textbf{Implementation Details.}
This section summarizes our implementation settings. The training process is executed on a distributed TensorFlow \cite{tensorflow2016abadi} platform, consisting of 10 parameter servers and 10 workers, each with 2 GPUs. To ensure a fair comparison with baselines, we configure the total number of transformer layers to 3, with one layer for the encoder and two layers for the decoder. The input dimension is 128 and the hidden size per transformer layer is 640. For MPQ, the hyperparameter \(\alpha\) used in the training process of MPQ is set to 0.001. Because MPQ training is not the core focus and these weights have minimal impact once MPQ converges, we do not perform a sensitivity study on them. For flexible reasoning paths, MPQ uses 8 codebooks with 300 codewords each. The tokenizers process both text and image features for the industrial and public datasets. Text embeddings are generated using the pre-trained Qwen3-Embedding 7B model \cite{qwen3embedding} and image embeddings are generated using the pre-trained Pailitao v8 model \cite{zhang2018visual}, both with a dimensionality of 256. We set the clipping parameter $\epsilon$ to 0.15 to balance exploration and exploitation. In MSRA, the number of future steps is $h=3$ and the time-decay coefficient is $w=0.8$. In CORP, the self-reflection threshold is $\theta=0.06$. All other hyperparameters of these modules have been carefully tuned through sensitivity analysis, as detailed in Section~\ref{sec:Sensitivity}.

\subsection{Experiment Performance (RQ1)}
Table \ref{table:offline_experiment_results} summarizes the overall performance of our REG4Rec against multiple baselines on both public and industrial datasets. The best results are highlighted in bold, while the second-best results are underlined. From these results, we make the following observations:
\begin{itemize}[noitemsep, topsep=0pt, leftmargin=*]
\item Our REG4Rec consistently outperforms all baseline methods across the aforementioned datasets, achieving average relative improvements of \textbf{10.06\%} in R@5, \textbf{7.30\%} in N@5, \textbf{11.01\%} in R@10 and \textbf{9.88\%} in N@10 compared to the best-performing baseline on each dataset.

\item $S^3$-Rec significantly outperforms SASRec on most datasets. This improvement demonstrates the effectiveness of the transformer-based architecture in capturing diverse patterns. Similarly, our REG4Rec captures richer intent patterns through the incorporation of MPQ and further improves performance.

\item TIGER introduces an end-to-end GR model that directly generates semantic IDs for the target item, outperforming traditional transformer-based baselines across all metrics and datasets. Building upon this foundation, COBRA expands the semantic space by integrating dense representations with sparse IDs through a cascading generation process. This approach enables COBRA to outperform TIGER on all datasets, achieving state-of-the-art results among baselines on several datasets.

\item As an extension of SASRec, ReaRec improves average performance by more than 13.13\% through an implicit multi‑step reasoning, but it lacks explicit reasoning process and its paths are semantically ambiguous. In contrast, STEAM achieves leading performance among all baselines on the Sports and Industrial datasets by using textually similar items as explicit reasoning steps and training with GRPO. However, the reasoning paths in STEAM are predetermined, which limits its ability to capture the diverse user interests. More fundamentally, existing methods also lack a mechanism to validate path reliability, undermining the quality of their recommendations. In contrast, REG4Rec constructs flexible and reliable reasoning paths, enabling more dynamic and accurate preference modeling and delivering consistent gains over both public and industrial baselines.
\end{itemize}

In summary, REG4Rec preserves the strengths of GR while enhancing reasoning with flexible and reliability‑aware mechanisms, overcoming the limitations of existing methods and achieving state‑of‑the‑art performance across all datasets.

\subsection{Ablation Study (RQ2)}
To assess the effectiveness of each component, we conduct ablation studies on the industrial dataset, which is the most complex and representative scenario on our recommendation platform. As shown in Table~\ref{fig:data_scale_comparison}, the industrial dataset is orders of magnitude larger than public datasets, providing a more realistic and challenging evaluation environment.

\begin{itemize}[noitemsep, topsep=0pt, leftmargin=*]
\item \textbf{w/o MPQ}: Replaces MPQ with the widely used RQ-VAE \cite{yang2025sparse,zhang2025slow}. To ensure a fair comparison, in RQ-VAE, both the number of codebooks and the size of each codebook are kept consistent with those in MPQ. Additionally, the same multimodal representations are used as input for both models. In RQ-VAE, since the token for the item is sequential, the codebook order is fixed at inference.
\item \textbf{w/o CRSS}: Disable dynamic ordering for parallel codebooks so codewords are generated in a fixed order.
\item \textbf{w/o PARS}: Remove the RL-based component for reasoning enhancement, reducing the model to a pre‑trained one with MPQ and MSRA.
\item \textbf{w/o MSRA}: Remove the multi-step reward strategy.
\item \textbf{w/o CORP}: Remove the consistency‑oriented reflection strategy so inconsistent paths are not filtered at inference.
\end{itemize}

Table \ref{table:ablation_experiment_results} presents the performance of these ablation experiments. We summarize the main observations below:
\begin{itemize}[noitemsep, topsep=0pt, leftmargin=*]
\item Replacing MPQ with RQ-VAE leads to an average performance drop of \textbf{5.03\%}, indicating that parallel tokenizers for the item are better suited to dynamic reasoning than the sequential RQ‑VAE, which has limited semantic expressiveness.

\item Removing CRSS causes an average drop of \textbf{2.84\%}, suggesting that the dynamic ordering mechanism in MPQ is crucial for effectiveness by expanding the exploration space.

\item PARS is our core contribution, as it incorporates reasoning ability into GR. Removing it leads to a significant average drop of \textbf{7.80\%}, which highlights the importance of guiding the model to select reasoning paths with high consistency.

\item Removing MSRA degrades all metrics, with larger drops on top‑10 (\textbf{5.73\%} on average) than top‑5 (\textbf{3.66\%} on average), indicating its role in capturing broader, more generalizable preferences.

\item Without CORP, there is a clear decline in model performance, with a more substantial decrease observed in the top-5 metrics (\textbf{5.73\%} on average) than in the top-10 metrics (\textbf{4.15\%} on average). This indicates that the reflection mechanism is particularly effective in capturing users' accurate interests.
\end{itemize}

\begin{table}[htbp!]
\caption{Ablation Study of REG4Rec.}
\renewcommand{\arraystretch}{1.3}
\centering
\resizebox{250pt}{!}{ 
\begin{tabular}{c|cccc}
\toprule
\multirow{1}{*}{\centering Method} 
 & R@5 & N@5 & R@10 & N@10 \\ 
 \midrule
\textbf{REG4Rec}
&\textbf{0.1094}&\textbf{0.0738}&\textbf{0.1569}&\textbf{0.0890}\\
w/o MPQ &0.1042&0.0690&0.1505&0.0838\\
w/o CRSS &0.1068&0.0706&0.1538&0.0857\\ 
w/o PARS &0.1010&0.0677&0.1455&0.0819\\ 
w/o MSRA &0.1059&0.0706&0.1487&0.0831\\ 
w/o CORP &0.1032&0.0695&0.1504&0.0853\\ 
\bottomrule
\end{tabular}
}
\label{table:ablation_experiment_results}
\end{table}

\begin{figure}[htbp]
  \includegraphics[width=0.48\textwidth]{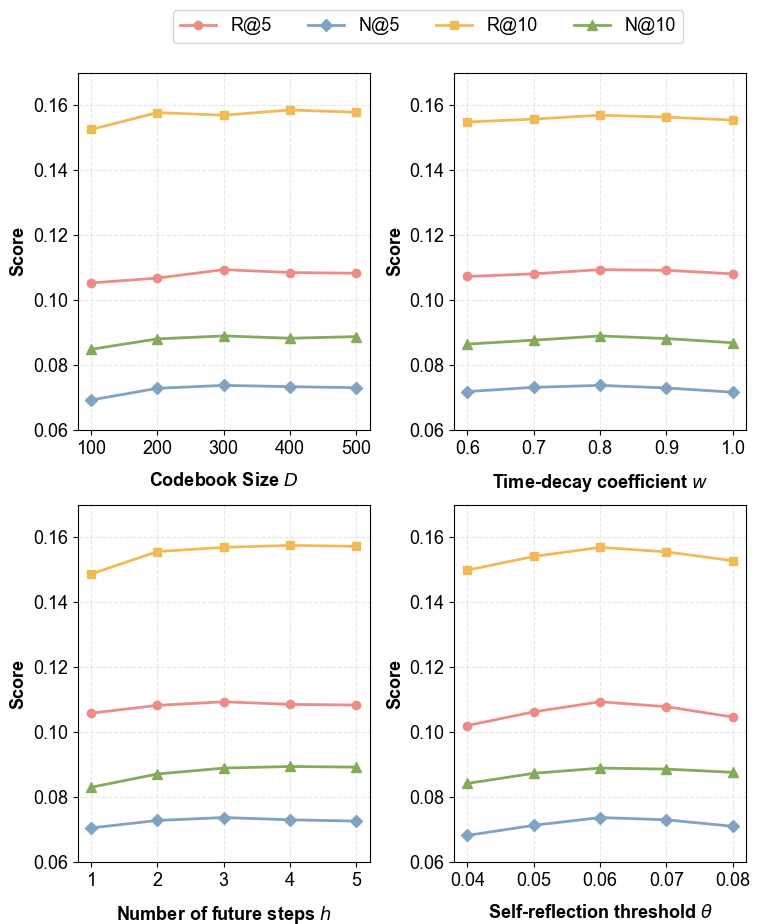}
  \caption{Sensitivity Analysis}
  \label{fig:sensitive_v1}
\end{figure}
\subsection{Sensitivity Analysis (RQ3)}
~\label{sec:Sensitivity}
In this section, we investigate the sensitivity of four hyperparameters: the number of codewords in each codebook \(D\), the time-decay coefficient \(\alpha\), the number of future steps \(h\) in MSRA and the self-reflection threshold \(\theta\) of CORP used for early path pruning during the reasoning process. We run all experiments on the industrial dataset, varying each hyperparameter over five values: \(D\) $\in$ (100, 200, 300, 400, 500), \(\alpha\) $\in$ (0.6, 0.7, 0.8, 0.9, 1.0), \(h\) $\in$ (1, 2, 3, 4, 5) and \(\theta\) $\in$ (0.04, 0.05, 0.06, 0.07, 0.08). Figure \ref{fig:sensitive_v1} illustrates the performance of these hyperparameter analysis experiments. We derive the following observations:
\begin{itemize}[noitemsep, topsep=0pt, leftmargin=*]
\item \textbf{Codebook Size \(D\)}: REG4Rec remains strong and stable as \(D\) increases from 200 to 500, with only a slight drop at \(D=100\), indicating that once the representation space is sufficiently large, performance is insensitive to $D$.
\item \textbf{Time-decay coefficient \(w\)}: As \(w\) decreases from 1.0 to 0.8, the performance of REG4Rec improves steadily (\textbf{+1.66\%} on average), confirming the effectiveness of the time decay strategy in MSRA. However, further reducing \(w\) slightly hurts performance, because over‑suppressing long‑term signals becomes harmful.
\item \textbf{Number of future steps \(h\)}: When \(h\) increases from 1 to 3, the performance of REG4Rec improves steadily, showing that incorporating a sufficient number of future behaviors is helpful for achieving optimal results. However, further increasing \(h\) yields diminishing returns, as the additional behavioral signals are increasingly contaminated by noise, eventually offsetting the benefits of the MSRA.
\item \textbf{Self-reflection threshold \(\theta\)}: As \(\theta\) decreases from 0.08, the consistency constraint becomes stricter. Model performance initially improves steadily and reaches its peak around \(\theta=0.06\) (\textbf{+3.07\%} on average). However, further reducing \(\theta\) leads to a decline in performance, indicating that overly strict constraints are detrimental. This suggests that a moderate level of consistency constraints is sufficient to enhance inference quality.
\end{itemize}

\subsection{Reasoning Step Scaling (RQ4)}
\begin{figure}[htbp]
  \includegraphics[width=0.48\textwidth]{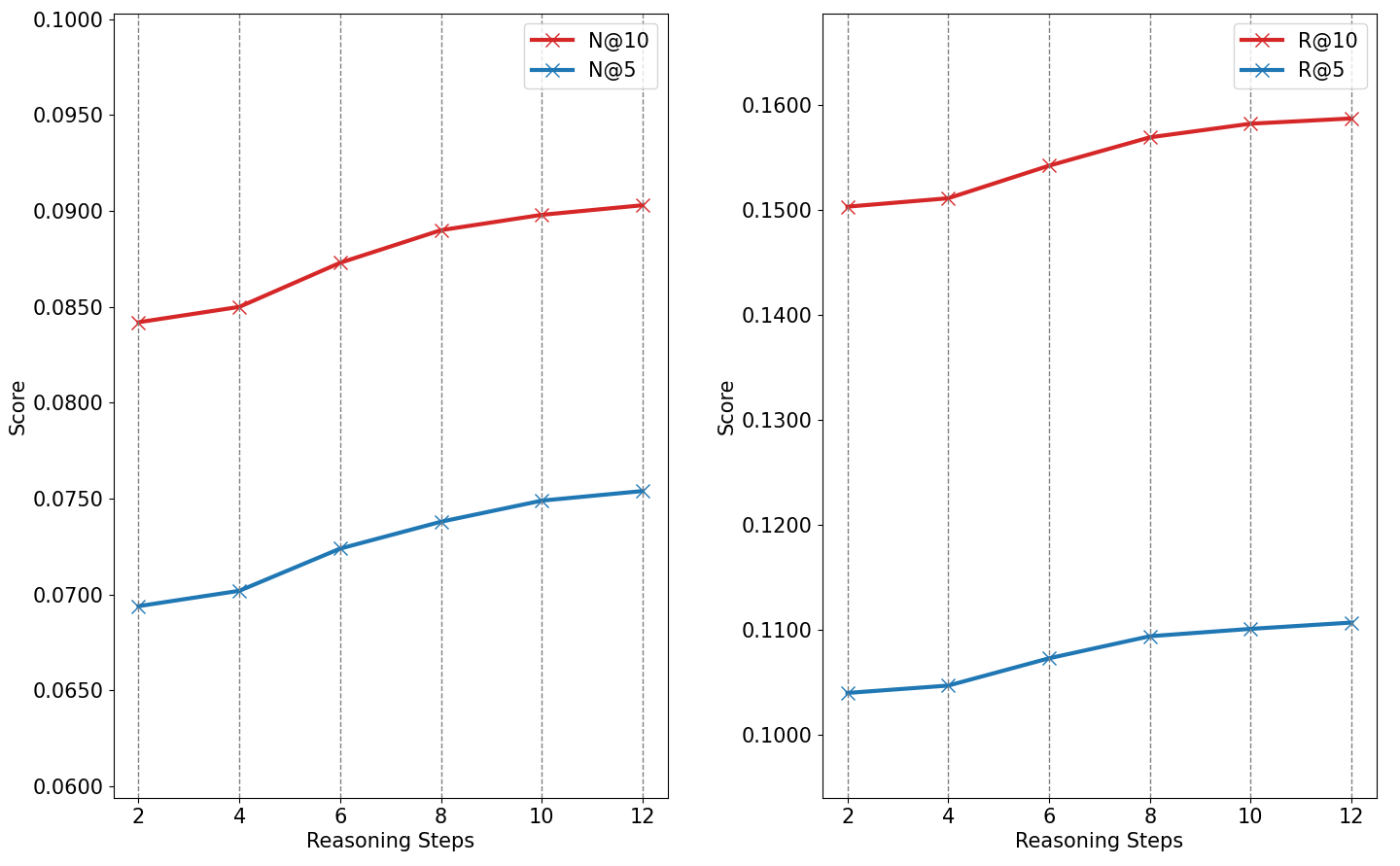}
  \caption{The performance of REG4Rec across different numbers of reasoning steps.}
  \label{fig:scaling}
\end{figure}

In this section, we evaluate how performance scales with the number of reasoning steps at inference. To this end, we vary the number of codebooks $M$, which determines the maximum number of reasoning steps. As shown in Figure \ref{fig:scaling}, consistently increasing the reasoning steps from 2 to 12 leads to monotonic gains in N@K and R@K. These results demonstrate that REG4Rec exhibits robust scaling properties with reasoning depth. REG4Rec effectively leverages longer reasoning chains to capture more refined preferences while maintaining both training efficiency and inference reliability.

\subsection{Efficient Training Performance (RQ5)}
We evaluate LADQ (Section ~\ref{sec:Efficient Training Strategy}) during training. Following prior work \cite{wang2018training}, we measure throughput as batch-samples per second (QPS) and report R@5 and N@5 for recommendation quality. As shown in Table \ref{Efficient_Training}, on the industrial dataset, LADQ reaches 4.5 QPS, representing a 32.3\% speedup over the fp32 baseline. This acceleration comes at a negligible cost to accuracy, with only a 1.3\% relative drop in R@5 and a 0.8\% drop in N@5. In contrast, uniform quantization schemes (bf16 and fp8) offer higher QPS but incur larger accuracy degradation. Overall, LADQ offers a superior trade-off, achieving substantial training acceleration with minimal performance degradation.
\begin{table}[htbp]
    \centering
    \caption{Training performance under different quantization schemes on the industrial dataset. The performance differences compared to the baseline (fp32) are highlighted in bold.}
    \renewcommand{\arraystretch}{1.3}
    \label{Efficient_Training}
    \resizebox{250pt}{!}{  
        \begin{tabular}{c|c|c|c}
            \toprule
             Method & QPS & R@5 & N@5 \\
            \midrule
            fp32 & 3.4 & 0.1108 & 0.0744 \\
            bf16 & 5 (\textbf{+47.1\%}) & 0.0723 (\textbf{-34.7\%}) & 0.0476 (\textbf{-36.0\%}) \\
            fp8 & 6.3 (\textbf{+85.3\%}) & 0.0467 (\textbf{-57.9\%}) & 0.0296 (\textbf{-60.2\%}) \\
            LADQ  & 4.5 (\textbf{+32.3\%}) & 0.1094 (\textbf{-1.3\%}) & 0.0738 (\textbf{-0.8\%}) \\
            \bottomrule
        \end{tabular}
    }
\end{table}

\section{Online Experiments (RQ6)}
To further validate our approach, we conducted an online A/B test on an advertising recommendation platform of Alibaba, a leading e-commerce company in Southeast Asia, from July 18 to 22, 2025. The control group is based on SASRec, whereas the experimental group employs our proposed REG4Rec. Both groups consisted of 15\% randomly selected users. Specifically, we observed \textbf{5.60\%} increase in the \textbf{Advertising Revenue}, \textbf{1.81\%} increase in the \textbf{Click-Through Rate (CTR)} and \textbf{3.29\%} increase in the \textbf{Gross Merchandise Volume (GMV)}. The results of the online experiment once again confirm the performance of our proposed REG4Rec.

\begin{table}[htbp]
    \centering
    \caption{Results of the online A/B experiment, with all performance gains being statistically significant at $p < 0.05$.}
    \renewcommand{\arraystretch}{1.2}
    \label{dataset2}
    \resizebox{250pt}{!}{  
        \begin{tabular}{c|ccc}
            \toprule
             Method & Advertising Revenue & CTR  & GMV  \\
            \midrule
            REG4Rec & +5.60\% & +1.80\% & +3.29\% \\
            \bottomrule
        \end{tabular}
    }
\end{table}

\section{Conclusion and Future Work}
In this paper, we propose REG4Rec, a reasoning-enhanced GR model that addresses limitations of existing GR methods. REG4Rec employs a MoE-based parallel quantized codebook to construct a large-scale diverse reasoning pathway space, enabling adaptation to heterogeneous user intents. During training, we adopt the preference alignment for reasoning and multi-step reward augmentation strategies to dynamically select reasoning paths with high consistency and reliability. In the inference phase, a consistency-oriented self-reflection pruning strategy is introduced to accelerate online reasoning, addressing online deployment challenges in large-scale recommendation systems. Extensive offline experiments on real‑world datasets and an online deployment in a commercial advertising system demonstrate the effectiveness and practical value of REG4Rec. In summary, REG4Rec provides a novel and scalable solution for generative recommendation in large-scale systems, offering valuable insights for future research in this field.

Nevertheless, several avenues remain unexplored. First, REG4Rec's reasoning quality still depends on reward functions crafted for specific recommendation scenarios; developing self-adaptive or learned reward mechanisms could enable broader applicability across domains without extensive manual tuning. Second, while our multi-path reasoning enriches diversity, its full potential for explainability, causal reasoning, and user-controllable recommendations remains underexplored. Third, although REG4Rec is optimized for large-scale recommendation, extending its reasoning paradigm to multimodal signals, cross-domain transfer, or cold-start environments could greatly expand its impact. Finally, integrating human feedback more deeply into the reasoning-reflection loop may further enhance both reliability and user trust. Addressing these aspects would not only strengthen REG4Rec's scalability and adaptability but also push the boundaries of reasoning-enhanced generative recommendation in real-world applications.

\section{Ethical Considerations}
Throughout this research, we have strictly followed established ethical guidelines and best practices in data handling, experimentation, and system deployment. All experiments were conducted using widely available open-source datasets, as well as an anonymized industrial dataset, in which every piece of user-related information was thoroughly removed, masked, or aggregated to ensure full compliance with privacy regulations and to protect individual identities. Special care was taken to prevent any possibility of re-identification, including the removal of indirect identifiers and sensitive behavioral traces.
From a security and safety perspective, we implemented checks to ensure that the recommendations generated by REG4Rec pose no risk to user well-being, digital safety, or platform integrity. We also verified that no outputs promote harmful, misleading, or unsafe content. These measures, together with our privacy-preserving practices, ensure REG4Rec contributes positively to both research and real-world applications.

\end{document}